\documentclass[letterpaper, 12pt]{article}
\usepackage{latexsym}
\usepackage{amsmath}
\usepackage{amssymb}
\usepackage[latin1]{inputenc}
\usepackage{graphicx}
\newcommand{\bbibitem}[1]{\bibitem{#1}\marginpar{#1}}

\def\Label#1{\label{#1}%
  \smash{\hbox to0pt{\raise1ex\hbox{\tiny[#1]}\hss}}}
\def\noLabels{\let\Label=\label}
\def\nobbibitem{\let\bbibitem=\bibitem}

\begin{document}
\setcounter{page}{0} \noLabels

\rightline{UPR-T-1175, hep-th/0701263}
\vskip 1cm

\centerline{\Large \bf Trace Anomaly in Geometric Discretization}
\vskip 1cm

\renewcommand{\thefootnote}{\fnsymbol{footnote}}
\centerline{{\bf Bart{\l}omiej
Czech${}$\footnote{czech@sas.upenn.edu}}} \vskip .5cm
\centerline{\it David Rittenhouse Laboratories, University of
Pennsylvania,} \centerline{\it Philadelphia, PA 19104, U.S.A.}

\vskip 1cm \setcounter{footnote}{0}
\renewcommand{\thefootnote}{\arabic{footnote}}

\begin{abstract}
I develop the simplest geometric-discretized analogue of two
dimensional scalar field theory, which qualitatively reproduces
the trace anomaly of the continuous theory. The discrete analogue
provides an interpretation of the trace anomaly in terms of a
non-trivial transformation of electric-magnetic duality-invariant
modes of resistor networks that accommodate both electric and
magnetic charge currents.
\end{abstract}

\newpage
\pagestyle{plain}

\section{Introduction}

Finding new formulations of familiar concepts frequently proves
beneficial in physics. This may be motivated by computational
needs, as in lattice field theory, or by a hope to gain new
fundamental insights. Since the 1960s the idea of geometric
discretization has lured physicists with promises of new ways of
looking at field theory and quantum gravity. After Regge calculus
\cite{regge}, a discretized theory of gravity, a discrete
understanding of Abelian Chern-Simons theory \cite{doubling}, $BF$
field theory \cite{topbf} and chiral fermions \cite{chiral} has
been attained. Not surprisingly, geometric discretization has been
most successful in modelling topological concepts and theories.
Meanwhile, progress in discretizing non-topological objects has
been less convincing. Although Desbrun et al. \cite{marsden}
provided explicit discrete analogues of geometric constructs
including the Hodge star and the Lie derivative, the complexity of
their formalism makes it difficult to derive insights from it. For
example, at present there is no discrete derivation of the
well-known trace anomaly in two dimensions.

Here I develop a new, simple discretization of scalar field theory
and use it to derive the form of the trace anomaly from first
principles. I treat the discrete theory as defined independently
of its continuous counterpart and work only with constructs which
are intrinsic to the discrete theory on a triangulation. The
starting point of the discretization is the identification of
$p$-forms on a manifold with $p$-cochains on a triangulation, and
thereafter the calculation proceeds along a path which is
essentially selected by consistency requirements. As such, the
present work exhibits the simplest discrete analogue of
two-dimensional scalar field theory on a metric space.
%With a view
%to simplicity, I do not attempt to tune the discretization in
%order to reproduce the coefficient present in the continuous
%anomaly.

At the end of the calculation a new way of thinking about the
trace anomaly emerges. The manifold (triangulation) is viewed as
an electric-magnetic network and conformal transformations are
found to act non-trivially on the relevant charge densities.
Appealing to the correspondence with the continuous case, the
final form of the discrete trace anomaly may be viewed as singling
out a preferred, smeared out notion of curvature on a finite
triangulation of a two-dimensional manifold.\footnote{For an
alternative treatment, see Ko and Ro{\v c}ek \cite{rocek}.} More
on this in the discussion, which follows the review of Sec.
\ref{review} and the calculations of Sec. \ref{calculation}.

\section{Review of the continuous theory}
\label{review}

I begin with a brief review of the scalar theory on a smooth
compact orientable two-dimensional manifold $M$ without boundary,
based on \cite{ziliao}. This will help to fix all relevant
conventions and provide a necessary reference point for setting up
a dictionary between the continuous and discrete constructs.

A manifold $M$ is endowed with the de Rham complex $\Omega^*(M)$
\begin{equation}
0 \quad \stackrel{i}{\rightarrow} \quad \Omega^0(M) \quad
\stackrel{d_0}{\rightarrow} \quad \Omega^1(M) \quad
\stackrel{d_1}{\rightarrow} \quad \Omega^2(M) \quad
\stackrel{d_2}{\rightarrow} \quad 0 , \Label{deRham}
\end{equation}
where the maps $d_r$ satisfy $d_{r+1} d_r = 0$, and the wedge
product
\begin{equation}
\begin{array}{rrcl}
\wedge^r\! : & \Omega^r(M) \ \times \ \Omega^{2-r}(M) &
\rightarrow & \mathbb{C} \\
& (\alpha^{(r)}, \, \gamma^{(2-r)}) & = & \int_M \alpha^{(r)}
\wedge^r \gamma^{(2-r)} = (-1)^r \, \int_M \gamma^{(2-r)}
\wedge^{2-r} \alpha^{(r)} .
\end{array}
\Label{wedgedef}
\end{equation}
The introduction of a metric $g_{\mu\nu}$ induces the Hodge star,
a natural isomorphism between $\Omega^r(M)$ and $\Omega^{2-r}(M)$,
which in turn induces a canonical inner product on $\Omega^r(M)$:
\begin{equation}
\begin{array}{rccl}
*_r : & \Omega^r(M) & \rightarrow &  \Omega^{2-r}(M) \\
\langle \ldots, \, \ldots\rangle^{(r)} : & \langle\alpha, \,
\beta\rangle & = & \int_M \alpha \wedge^r *\beta \, ,
\end{array}
\Label{stardef}
\end{equation}
where in the second line I suppressed the superscripts indicating
the degrees of the forms. I take the manifold $M$ to be
Riemannian, so the Hodge star satisfies:
\begin{equation}
*_{2-r} \, \circ \, *_r = (-1)^r \, .
\end{equation}
The Hodge star allows one to introduce operators $\delta_{r+1}$
adjoint to the exterior derivatives $d_{r}$:
\begin{equation}
\begin{array}{rccl}
\delta_{r+1} : & \Omega^{r+1}(M) & \rightarrow &  \Omega^r(M) \\
& \delta_{r+1} & = & *_{2-r} \, \circ \, d_{1-r} \, \circ \,
*_{r+1} \, .
\end{array}
\end{equation}
The resulting structures, the de Rham complex together with its
adjoint, are summarized in the diagram below:
\begin{equation}
0 \,\quad
{\begin{subarray}{c} i \\ \substack{\longrightarrow \\ \longleftarrow} \\
\delta_0 \end{subarray}} \,\quad \Omega^0(M) \,\quad
{\begin{subarray}{c} d_0 \\ \substack{\longrightarrow \\ \longleftarrow} \\
\delta_1 \end{subarray}} \,\quad \Omega^1(M) \,\quad
{\begin{subarray}{c} d_1 \\ \substack{\longrightarrow \\ \longleftarrow} \\
\delta_2 \end{subarray}} \,\quad \Omega^2(M) \,\quad
{\begin{subarray}{c} d_2 \\ \substack{\longrightarrow \\
\longleftarrow} \\ i \end{subarray}} \,\quad 0 \, .
\Label{codeRham}
\end{equation}

A 0-form field $\Phi$ coupled to a gravitational background on a
2-dimensional manifold $M$ without boundary has the action:
\begin{equation}
S = \frac{1}{2} \int_M d_0\Phi \wedge^1 *_1 d_0\Phi = \frac{1}{2}
\langle d_0\Phi, d_0\Phi\rangle^{(1)} \, . \Label{action}
\end{equation}
The appearance of $*_1$ in (\ref{action}) breeds a dependence on
$g_{\mu\nu}$ and is the ultimate reason for the theory not being
topological. The effective action
\begin{equation}
W[g] = -\log{\int [d\Phi] \, e^{-S[\Phi,g]}} \Label{effaction}
\end{equation}
suffers from an anomaly:
\begin{equation}
g_{\mu\nu}(\boldsymbol{x}) \frac{\delta W[g]}{\delta
g_{\mu\nu}(\boldsymbol{x})} = -\frac{1}{2} \sqrt{g}\,
\lim_{\epsilon \rightarrow 0} \, \langle \boldsymbol{x} |
e^{\epsilon \Delta} | \boldsymbol{x} \rangle \, .
\Label{contlocal}
\end{equation}
The small time diagonal heat kernel takes the form
\begin{equation}
\lim_{\epsilon \rightarrow 0} \,\langle \boldsymbol{x} |
e^{\epsilon \Delta} | \boldsymbol{x} \rangle =
\frac{1}{4\pi\epsilon}+\frac{1}{24\pi}R(\boldsymbol{x}) +
\mathcal{O}(\epsilon) \, ,
\end{equation}
which after the addition of a local counterterm $S\rightarrow S
-A/8\pi\epsilon$ yields
\begin{equation}
g_{\mu\nu}(\boldsymbol{x}) \frac{\delta W[g]}{\delta
g_{\mu\nu}(\boldsymbol{x})} =
 -\frac{1}{48 \pi} \sqrt{g} R(\boldsymbol{x}) \, .
 \Label{curvaturecont}
\end{equation}
The anomaly, integrated over $M$, gives:
\begin{equation}
\int_M g_{\mu\nu} \frac{\delta W[g]}{\delta g_{\mu\nu}} =
\frac{1}{6} \chi. \Label{contglobal}
\end{equation}

\section{The discrete theory}
\label{calculation}

I now set out to find the analogues of
eqs.(\ref{action}-\ref{contglobal}) for a theory defined on a
finite triangulation $\mathcal{G}$ of the manifold $M$.
$\mathcal{G}$ is most easily thought of as a graph $G(V,E)$
containing $|V|$ vertices and $|E|$ edges, embedded in $M$ in such
a way that no two edges cross. For technical convenience I shall
assume that $G$ does not contain loops and the number of edges $E$
is even. The embedding partitions $M$ into $|F|$ faces, which are
(homeomorphic to) triangles when $G$ corresponds to a
triangulation but may be (homeomorphic to) arbitrary polygons for
a more general $G$. The number of faces is such that the Euler
characteristics agree:
\begin{equation}
\chi_G = |V| + |F| - |E| = \chi_M\, .
\end{equation}
It is natural to associate $p$-forms on $M$ with $p$-cochains of
$\mathcal{G}$. Because we are interested in modelling a real
scalar $\Phi$, I shall take $\Omega^0_{\boldsymbol{x} \in M}
\rightarrow \Omega^0_{v \in V} \equiv \mathbb{R}$. Thus:
\begin{eqnarray}
\Omega^0(M) \longleftrightarrow \mathbb{R}^V \nonumber \\
\Omega^1(M) \longleftrightarrow \mathbb{R}^E \Label{corromega} \\
\Omega^2(M) \longleftrightarrow \mathbb{R}^F \nonumber
\end{eqnarray}
This set-up suggests a canonical correspondence between the
exterior derivative on $M$ and the co-boundary operator on
$\mathcal{G}$. To define the latter, assign to each edge $e \in E$
an arbitrary orientation, that is to say, think of one of the two
vertices incident to $e$ as ``initial'' and of the other one as
``final''. Then the $|E|\times |V|$ directed incidence matrix
$\mathbf{M}^G$ given by
\begin{equation}
\mathbf{M}^G_{ev} = \left\{ \begin{array}{ll}
 -1 & \textrm{if $v = \textrm{init}(e)$} \\
  1 & \textrm{if $v = \textrm{fin}(e)$}  \\
  0 & \textrm{otherwise} \end{array} \right. \Label{d0def}
\end{equation}
acts on $\mathbb{R}^V$ as a coboundary operator \cite{stanley} on
the space of 0-cochains:
\begin{equation}
\mathbf{M}^G\!: \,\, \mathbb{R}^V \rightarrow \mathbb{R}^E \quad
\longleftrightarrow \quad d_0\!: \,\, \Omega^0(M) \rightarrow
\Omega^1(M)
\Label{d0corr}
\end{equation}
The image of $\mathbf{M}^G$ in $\mathbb{R}^E$ is called the bond
space of $G$, denoted $\mathcal{B}(G)$. If we think of $G$ as
specifying an electrical network, $\mathcal{B}(G)$ is the space of
all possible potential differences. The orthogonal complement of
$\mathcal{B}(G)$ in $\mathbb{R}^E$ is the cycle space
$\mathcal{C}(G)$, the space of all circular flows in $G$. The
analogy with an electrical network will be useful to us later.

If $G$ represents a triangulation, the correspondence
(\ref{d0corr}) is canonically extended to $d_1$ as follows. Denote
every edge $e \in E$ by $[ij]$, where $v_i = \textrm{init}(e)$ and
$v_j = \textrm{fin}(e)$. In a similar way, denote every face $f
\in F$ by $[ijk]$, where $v_i,\, v_j,\, v_k$ are incident to $f$
and their ordering agrees with the orientation of the underlying
manifold $M$. Permutations act on $[ij],\, [ijk]$ like parity,
e.g. $[ij]=-[ji]$ and $[ijk]=[jki]=-[jik]$. We define the linear
operator $\mathbf{N}^G\!: \mathbb{R}^E \rightarrow \mathbb{R}^F$
via:
\begin{equation}
(\mathbf{N}^G \omega) ([ijk]) = \omega([ij]) + \omega([jk]) +
\omega([ki])\qquad\textrm{for } \omega \in \mathbb{R}^E
\Label{d1def}
\end{equation}
Then by construction $\mathbf{N}^G \mathbf{M}^G = 0$, which mimics
$d_1 d_0 = 0$ under the correspondence:
\begin{equation}
\mathbf{N}^G\!: \,\, \mathbb{R}^E \rightarrow \mathbb{R}^F \quad
\longleftrightarrow \quad d_1\!: \,\, \Omega^1(M) \rightarrow
\Omega^2(M)\, .
\Label{d1corr}
\end{equation}
It is easy to see that the image of $\mathbf{N}^G$ spans all of
$\mathbb{R}^F$ except the constant mode. This agrees with the
Hodge decomposition theorem, which stipulates that
\begin{equation}
\dim{\Omega^2(M)} = \dim{d_1 \Omega^1(M)} + b^2\, .
\end{equation}
The only remaining non-trivial consistency condition is the
antisymmetry of the wedge product $\wedge^1$, which is made
possible by the evenness of $|E|$. Consider any matrix of wedge
products $\Lambda^1(G)$:
\begin{equation}
\Lambda^1(G)_{\omega_1\omega_2} = (\omega_1, \, \omega_2)
\end{equation}
Then the analogous $\Lambda^0(G)$ is determined by enforcing
Stokes' theorem:
\begin{equation}
\int_M d_0 \alpha^{(0)} \wedge^1 \gamma^{(1)} = -\int_M
\alpha^{(0)} \wedge^0\, d_1\!\!\; \gamma^{(1)} \quad
\longleftrightarrow \quad (\mathbf{M}^G)^{\mathrm{T}} \Lambda^1(G)
= -\Lambda^0(G) \, \mathbf{N}^G . \Label{stokes}
\end{equation}
The above definition is supplemented by requiring that the matrix
element of $\Lambda^0(G)$ evaluated in the constant modes
$\mathbb{R}^F \setminus \textrm{Im}(\mathbf{N}^G)$ and
$\textrm{ker}(\mathbf{M}^G)$ be non-zero, so as to ensure
non-vanishing area of the manifold. In summary, the topological
data of the manifold $M$ find well-defined analogues in the
triangulation $\mathcal{G}$.

However, the introduction of a metric interrupts this complacent
state of affairs. The isomorphism (\ref{stardef}) requires
$\Omega^r(M)$ and $\Omega^{2-r}(M)$ to be of the same dimension.
Under the correspondence (\ref{corromega}) this translates into
$|V|=|F|$ and the only triangulation which satisfies this is the
tetrahedron triangulating a sphere. If we relax the condition of
working with a triangulation, the map (\ref{d1corr}) will be lost;
meanwhile, $|V|=|F|$ remains a cumbersome and non-trivial
condition.\footnote{This program was pursued in \cite{cubic}.} A
more general possibility is to replace $G$ with the union of the
graph and its dual, $G\cup \widetilde{G}$, with the understanding
that the resulting $p$-cochains should correspond to a
reduplicated space $\Omega^p(M)\oplus\Omega^p(M)$ \cite{doubling}.
In this paper I follow the latter strategy but abstain from the
onerous detail.

Consider two copies of the scalar field theory defined in
(\ref{action}). The fundamental scalars $\Phi_1$ and $\Phi_2$ are
naturally associated with the 0-cochains of $G\cup \widetilde{G}$
whose space is isomorphic to $\mathbb{R}^{V\oplus\widetilde{F}}$.
The previous notion of the exterior derivative $d_0
\leftrightarrow \mathbf{M}^G$ naturally extends to
$\widetilde{G}$. First one identifies the sets $E$ and
$\widetilde{E}$, that is, $\mathbb{R}^{\widetilde{E}} \ni
\hat{\tilde{e}} = \hat{e} \in \mathbb{R}^E$ and
$\mathbb{R}^{\widetilde{E}} = \mathbb{R}^E$. Then one extends the
orientation on $G$ to $\widetilde{G}$ in the following fashion:
for an edge $e \in G$ separating faces $f_i,\, f_j \in F$ we say
that $\textrm{init}(\tilde{e}) = \tilde{f}_i$ if in crossing from
$f_i$ to $f_j$ in $G$ one passes $\textrm{fin}(e)$ on the right
and $\textrm{init}(e)$ on the left; otherwise $\tilde{f}_i =
\textrm{fin}(\tilde{e})$. This prescription is well-defined by
virtue of the orientability of the manifold $M$ in which $G$ is
embedded. Then one forms an $|\widetilde{E}| \times
|\widetilde{F}|$ (i.e. $|E|\times |F|$) directed incidence matrix
$\mathbf{M}^{\widetilde{G}}$ as in (\ref{d0def}):
\begin{equation}
\mathbf{M}^{\widetilde{G}}_{e\tilde{f}} = \left\{
\begin{array}{ll}
 -1 & \textrm{if $\tilde{f} = \textrm{init}(\tilde{e})$} \\
  1 & \textrm{if $\tilde{f} = \textrm{fin}(\tilde{e})$}  \\
  0 & \textrm{otherwise} \end{array} \right.
\end{equation}
The matrices $\mathbf{M}^G$ and $\mathbf{M}^{\widetilde{G}}$ can
be concatenated to form the collective incidence matrix
$\mathbf{M}^{G\cup \widetilde{G}}$, whose dimensions are $|E|
\times (|V|+|F|)$. It is easy to see that
$\mathbf{M}^{\widetilde{G}}$ sends $\mathbb{R}^{\widetilde{F}}$
into $\mathcal{C}(G)$, that is, $\mathcal{B}(G) \! \perp \!
\mathcal{B}(\widetilde{G})$. In other words, a potential
difference in the network $G$ is necessarily a circular flow in
the network $\widetilde{G}$. However, the converse is not true for
non-zero genera of $M$. There are $2g$ linearly independent flows
which are circular in both $G$ and $\widetilde{G}$, i.e.
$\dim{\mathcal{C}(G)\cap\mathcal{C}(\widetilde{G})}=2g$. This
observation will be of essential importance later.

Because we are working in $G\cup \widetilde{G}$, the triangular
character of the underlying graph is in general lost and we do not
have the luxury of extending (\ref{d1def}) to the present case.
However, since the action for two scalar fields reads
\begin{equation}
S_{12} = \frac{1}{2} \int_M d\Phi_1 \wedge *d\Phi_1 + \frac{1}{2}
\int_M d\Phi_2 \wedge *d\Phi_2 = \frac{1}{2} \langle d\Phi_1,\,
d\Phi_1\rangle + \frac{1}{2} \langle d\Phi_2,\, d\Phi_2\rangle\, ,
\Label{action2}
\end{equation}
finding an explicit operator representing $d_1$ is not necessary.
Similarly, an explicit representation of $\wedge^1$ is redundant
as the only combination that shows up in the action is $\wedge^1
*_1$. Thus, in the following I shall not be concerned with
searching for discrete analogues of all the standard machinery of
algebraic topology and geometry. Instead, I shall limit myself to
making a simple ansatz for the discrete representation of $\int_M
\ldots \wedge^1 *_1 \ldots = \langle
\ldots,\,\ldots\rangle^{(1)}$. This, complemented with a notion of
area and the correspondence $\mathbf{M}^{G\cup\widetilde{G}}
\leftrightarrow d_0$, will be sufficient for determining the trace
anomaly.

The introduction of a metric is in order. This is done by
assigning to each edge $e \in E$ a length $g_e$. I then take the
inner product on $\mathbb{R}^E$ to be:
\begin{eqnarray}
\mathbf{J}\phantom{_{ee'}} & \leftrightarrow & \langle
\ldots,\,\ldots\rangle^{(1)} \nonumber \\
\mathbf{J}_{ee'} & = & g_e \, \delta_{ee'} \, \Label{definnerpr}
\end{eqnarray}
To complete the mapping, a notion of area is necessary. This is
defined as:
\begin{equation}
A = \langle 1,\, 1\rangle^{(0)} = \int_M \sqrt{g} \leftrightarrow
\sum_{e \in E} g_e = \frac{1}{2} \sum_{v \in V} d_v = \frac{1}{2}
\sum_{\tilde{f} \in \widetilde{F}} d_{\tilde{f}} = \textrm{vol}(G)
\, , \Label{areacorr}
\end{equation}
where the degree of each vertex $d_v$ is the sum of the lengths of
the edges incident to $v$. The correspondence (\ref{areacorr}) has
been considered in mathematical literature (viz. Sec. 3.4 in
\cite{chung}), although it is not unique. Because differences
between the present approach and others (for example,
\cite{rocek}) may ultimately be traced to the choice of metric,
eqs. (\ref{definnerpr}-\ref{areacorr}) represent a key step in the
present development. I shall return to this point in the
discussion.

The correspondence (\ref{definnerpr}-\ref{areacorr}) is
well-motivated. From an abstract point of view, no prior
restrictions constrain the discrete analogue of the Hodge star so
the initial choice of mapping $A \leftrightarrow \textrm{vol}(G)$
is unrestrained. After that, the inner product $\langle\ldots,\,
\ldots\rangle^{(1)}$ is expected to have the same scaling as
$\langle \ldots,\, \ldots \rangle^{(0)}$ and exhibit an
appropriate metric-related notion of locality in $\mathbb{R}^E$.
This essentially determines $\langle\ldots,\, \ldots\rangle^{(1)}$
in a unique way.

On a more intuitive level, the analogy with an electrical network
is helpful. The zero modes in $\mathbb{R}^V$ and
$\mathbb{R}^{\widetilde{F}}$, the respective constants, can be
thought of as specifying constant densities of excess $\Phi_1$ and
$\Phi_2$ charge in the network. If we envision the charges as
residing along the wires of the network, then it is a canonical
choice to demand:
\begin{equation}
Q_{total} = \sigma_0 \, A \leftrightarrow \lambda_0 \, \sum_{e \in
E} g_e = Q_{total}\, ,
\end{equation}
where $\sigma_0$ and $\lambda_0$ denote the respective constant
charge densities. The vectors in $\mathbb{R}^E$ specify flows in
the network. Since $\langle 1,\, 1\rangle^{(0)} \propto
Q_{total}$, it is natural to require that $\langle 1_e,\,
1_e\rangle^{(1)}$ be proportional to the amount charge flowing in
$e \in E$, which in turn is proportional to the length of the edge
$g_e$. This motivates (\ref{definnerpr}).

With the definitions of $\mathbf{M}^{G\cup \widetilde{G}}
\leftrightarrow d_0$ and $\langle \ldots,\, \ldots\rangle^{(1)}$
in place, the action (\ref{action2}) reads:
\begin{equation}
S_{12} = \frac{1}{2} \Psi^{\mathrm{T}} (\mathbf{M}^{G\cup
\widetilde{G}})^{\mathrm{T}} \, \mathbf{J} \, \mathbf{M}^{G\cup
\widetilde{G}} \Psi\, , \Label{actiondiscr}
\end{equation}
where $\Psi \in \mathbb{R}^{V \oplus \widetilde{F}}$ represents
both $\Phi_1$ and $\Phi_2$ and lives in the direct sum of their
respective field spaces. The effective action is given by
\begin{equation}
W_{12}[g] = -\log{\int [d\Psi] e^{-S_{12}}}\, .
\Label{effactiondiscr}
\end{equation}

There are two zero modes which need to be regulated, one for each
of $\Phi_1,\, \Phi_2$. They are given by the constant modes on $V$
and $\widetilde{F}$, respectively. Using the electrical networks
analogy, it is natural to pick the following regularization
scheme. Suppose the network is capable of sustaining static charge
densities only up to a maximal value $\lambda_{max}^i, \,\, i=0,\,
1$. In terms of the normalized zero mode $\psi_0^i =
(\textrm{vol}(G))^{-1/2}$, the charge density is given by $0 \leq
\lambda_0^i = c_0^i \psi_0^i = c_0^i ( \textrm{vol}(G))^{-1/2}
\leq \lambda_{max}^i$. Then the integral over each zero mode
returns the value of $(c_0^i)_{max} = \lambda_{max}^i
(\textrm{vol}(G))^{1/2}$. Thus, eq. (\ref{effactiondiscr}) yields
\begin{equation}
W_{12}[g] = -\log{\textrm{vol}(G)} + \log{\det{' \left\{
(\mathbf{M}^{G\cup \widetilde{G}})^{\mathrm{T}} \, \mathbf{J} \,
\mathbf{M}^{G\cup \widetilde{G}} \right\} }} \Label{effactiontr}
\end{equation}
up to constant terms involving $\lambda_{max}^i$.

The second term in (\ref{effactiontr}) cannot be split into
logarithms of determinants of the factor matrices because the rank
of $\mathbf{M}^{G\cup \widetilde{G}}$ is lower than that of
$\mathbf{J}$. The obstruction to doing so lies, therefore, in the
zero modes of $(\mathbf{M}^{G\cup \widetilde{G}})^{\mathrm{T}}:
\mathbb{R}^E \rightarrow (\mathbb{R}^{V\oplus
\widetilde{F}})^{\vee}$. I proceed to regularize these modes.

The strategy will be analogous to the regularization of the zero
modes of $\mathbf{M}^{G\cup \widetilde{G}}$. The relevant zero
modes are flows which are circular both with respect to $G$ and
$\widetilde{G}$. There are $|E|-(|V|+|F|-2) = 2-\chi$ linearly
independent such flows. Suppose that for a particular zero mode
$\tau_i \in \mathcal{C}(G) \cap \mathcal{C}(\widetilde{G})$ the
underlying electrical network can only support charge fluxes not
exceeding a maximal value $\mathcal{I}_i^{max}$. This means that
the flux density over an edge $e$ is bounded by:
\begin{equation}
0 \leq c_i \, \tau_i(e)\, \Big( \sum_{e \in E} \tau_i(e)^2 g_e
\Big)^{-1/2} \leq \, \mathcal{I}_i^{max} \tau_i(e)
\end{equation}
and the volume of integration over $c_i$ is $\big(\sum_{e \in E}
\tau_i(e)^2 g_e\big)^{1/2}\, \mathcal{I}_i^{max}$. The
regularization allows us to treat the determinant in
(\ref{effactiontr}) as if it were non-singular:
\begin{equation}
\det{' \left\{ (\mathbf{M}^{G\cup \widetilde{G}})^{\mathrm{T}} \,
\mathbf{J} \, \mathbf{M}^{G\cup \widetilde{G}} \right\} } =
\det{\mathbf{J}} \, \det{\left\{\mathbf{M}^{G\cup \widetilde{G}}
(\mathbf{M}^{G\cup \widetilde{G}})^{\mathrm{T}}\right\} }\, .
\end{equation}
The second determinant is not primed as it includes the
regularized zero mode contributions. We have:
\begin{equation}
\det{\left\{\mathbf{M}^{G\cup \widetilde{G}} (\mathbf{M}^{G\cup
\widetilde{G}})^{\mathrm{T}}\right\} } = \det{'
\left\{\mathbf{M}^{G\cup \widetilde{G}} (\mathbf{M}^{G\cup
\widetilde{G}})^{\mathrm{T}}\right\} } \, \prod_{i=1}^{2-\chi}
\mathcal{I}_i^{max} \Big( \sum_{e \in E} \tau_i(e)^2 g_e
\Big)^{1/2}
\end{equation}
The index $i$ runs over an orthonormal basis $\tau_i$ of
$\mathcal{C}(G)\cap \mathcal{C}(\widehat{G})$.

Although the first factor is independent of the metric and will
merely contribute a constant term to $W[g]$, it is amusing to note
that it is blessed with a beautiful combinatorial interpretation.
If neither $G$ nor $\widetilde{G}$ contains loops, then the
determinant factorizes over $G,\,\widetilde{G}$ and according to
the matrix tree theorem \cite{kirchhoff}
\begin{equation}
\det{' \left\{\mathbf{M}^{G} (\mathbf{M}^{G})^{\mathrm{T}}\right\}
} = |E|\,\, \#\! \big\{ \textrm{spanning trees of }G \big\}
\end{equation}
with the analogous equality holding for $\widetilde{G}$. This
result, first demonstrated by Kirchhoff in 1847, is useful in
calculating total resistances in electric networks which are
neither parallel nor in series.

Putting all the results together and dropping constant terms, the
effective action takes the form
\begin{equation}
W[g] = -\frac{1}{2}\log{\Big( \sum_{e \in E} g_e \Big) }
+\frac{1}{4} \sum_{i=1}^{2-\chi} \log{\Big( \sum_{e \in E}
\tau_i(e)^2 g_e \Big) } + \frac{1}{2}\sum_{e \in E} \log{g_e} \, .
\Label{effactionfin}
\end{equation}
Here I have restored the factor of $1/2$ coming from the fact that
$W_{12}$ contains contributions from two scalars $\Phi_1, \,
\Phi_2$. The conformal variation of $W$ is given by
\begin{equation}
g_e \frac{\delta W[g]}{\delta g_e} = -\frac{1}{2}
\frac{g_e}{\sum_{e \in E} g_e} + \frac{1}{4} \sum_{i=1}^{2-\chi}
\frac{\tau_i(e)^2 g_e}{\sum_{e \in E} \tau_i(e)^2 g_e} +
\frac{1}{2} \, .\Label{curvaturediscr0}
\end{equation}
As in the continuous case, the last term may be cancelled by
introducing a local $\Phi$-independent counterterm $S \rightarrow
S - 1/2\,\, \textrm{Tr}\log{\mathbf{J}}$. The first two terms,
however, form a topological invariant:
\begin{equation}
\sum_{e \in E} g_e \frac{\delta W[g]}{\delta g_e} = -\frac{1}{4}
\chi\, .\Label{anomdiscr}
\end{equation}
The correspondence with the discrete result (\ref{curvaturecont},
\ref{contglobal}) is imperfect - the coefficients do not agree.
Nevertheless, eq. (\ref{curvaturediscr0}) singles out a preferred
discretized notion of curvature, distinct from that of
\cite{regge}:
\begin{equation}
\sqrt{g} \, R(\boldsymbol{x}) \quad
\stackrel{\propto}{\longleftrightarrow} \quad g_e \,\,\!
\Big(\frac{1}{2} \sum_{i=1}^{2-\chi} \frac{\tau_i(e)^2}{\sum_{e
\in E} \tau_i(e)^2 g_e} - \frac{1}{\sum_{e \in E} g_e} \Big) \, .
\Label{proposal}
\end{equation}

\section{Discussion}

This paper presents the shortest route to discussing the trace
anomaly in discretized settings. As a starting point, the
canonical correspondence between $p$-forms and $p$-cochains on a
triangulation was adopted, along with the natural definition of
the $0^{\rm th}$ exterior derivative. After that, the metric was
introduced via the Hodge star, leading to a doubling of the
field space for consistency. Every construct introduced in this
development is strictly necessary for a discussion of the trace
anomaly, and conversely, the construction is free of redundancies.
It is in this sense that the presented theory is the simplest
possible in the equivalence class of consistent discrete
two-dimensional scalar field theories exhibiting trace anomalies.

The derivation involved one arbitrary choice - that of the
discrete Hodge star $*_1$ (and therefore the metric) in eqs.
(\ref{definnerpr}-\ref{areacorr}). The correspondence adopted in
this paper was motivated by an effort to treat the objects
intrinsic to the triangulation - vertices, edges and edge lengths
- as fundamental, and avoid making recourse to an explicit
embedding.\footnote{In particular, the theory presented herein is
consistent even if the edge lengths defining the underlying
discrete space do not satisfy the triangle inequality.} It should
be borne in mind, however, that other equally motivated choices
are possible. Of these, a notable one is that employed by Ko and
Ro{\v c}ek in a related paper \cite{rocek}, which follows the
tenets of Regge calculus. In it, the discrete theory is implicitly
assumed to live on a piecewise linear space consisting of a union
of flat triangles, so that the area of the manifold does not
conform with eq. (\ref{areacorr}). This leads to an effective
action, whose conformal variation features the conical curvature
and which differs from eq. (\ref{effactionfin}). However, the
general approach presented in this work is robust. With a different input metric to replace eqs.
(\ref{definnerpr}-\ref{areacorr}), it should in principle reproduce
the result of \cite{rocek}. The numeric disagreement between the
discrete anomaly (\ref{anomdiscr}) and its continuous counterpart
(\ref{contglobal}) is likewise easily fixed by a trivial
redefinition of the metric $g_e \rightarrow g_e^{-3/2}$ in eq.
(\ref{definnerpr}). It would be interesting to understand the
exponent $-3/2$ in detail, as it provides a quantitative bridge
between two-dimensional scalar theory and electric-magnetic
resistor networks.

The latter analogy is a nice bonus of the calculation presented in
Sec. \ref{calculation}. We interpret the zero modes of
$\mathbf{M}$, $\psi_0^i$ as charge densities and the zero modes of
$\mathbf{M}^{\textrm{T}}$, $\tau_i$'s as circular flows in a
network of resistors. The network is electro-magnetic as both
electric and magnetic charges propagate in it. This is reflected
in the fact that the field $\Psi$ has two independent zero modes,
one living on $G$ and the other on $\widetilde{G}$. Indeed, upon
the exchange $G\rightleftharpoons \widetilde{G}$ we have:
\begin{equation}
\begin{array}{rcl} \mathcal{B}(G) & \hookrightarrow &
\mathcal{C}(\widetilde{G}) \\
\mathcal{B}(\widetilde{G}) & \hookrightarrow & \mathcal{C}(G) \, .
\end{array}
\end{equation}
This operation is easily seen as an interchange of the
two-dimensional divergence and curl, which effects an
electric-magnetic duality. For $g>0$ the network contains flows
which are cyclic both from the electric and the magnetic points of
view, given by $\mathcal{C}(\widetilde{G})\cap\mathcal{C}(G)$.
Populating such a circular flow diminishes the ability of the
network to respond to differences in potentials because it takes
away from the pool of propagating charges. Hence the contribution
of the modes $\tau_i \in
\mathcal{C}(\widetilde{G})\cap\mathcal{C}(G)$ to the effective
action is opposite from that of the zero modes of the fundamental
field $\psi_0^i$, which correspond to free excess charges. As
$\tau_i, \, \psi_0^i$ are zero modes, this effect is not countered
by a corresponding rescaling of the Laplacian eigenvalues, as is
the case for all other modes. As a result, $\tau_i, \, \psi_0^i$
entirely capture the effect of conformal transformations on the
effective action. This provides a natural interpretation of the
familiar result
\begin{equation}
\int_M g_{\mu\nu} \frac{\delta W[g]}{\delta g_{\mu\nu}} \ \propto
\ \textrm{ind }\!D \qquad\qquad \textrm{with }D = d_1 + \delta_1
\, , \Label{indexd}
\end{equation}
where $D$ acts on the space of 1-forms $\Omega^1(M)$. In this
language, the need for a doubling of the degrees of freedom,
which is ubiquitous in geometric discretization \cite{doubling},
is understood as arising from the requirement that 
electric and magnetic degrees of freedom be treated in a 
duality-symmetric way. Geometric discretization
without doubling treats the two kinds of degrees of freedom
asymmetrically, viz. the paragraph below equation (\ref{stokes}), 
which in itself is not inconsistent.
Invariance under the electric-magnetic duality becomes necessary, 
however, after the introduction of the metric and the Hodge star.
It is satisfying to have
geometric discretization provide an intuitive understanding of the
trace anomaly in both physical and algebro-geometric terms.

The route toward discretization followed in Sec. \ref{calculation}
may also provide attractive settings for modelling emergent
locality. The discrete counterpart of curvature proposed in
(\ref{proposal}) is not local in that it depends on the detailed
form of the zero modes, which is determined by the full structure
of the triangulation. This is again to be contrasted with the
approach of Ko and Ro{\v c}ek \cite{rocek}, who exhibited the
effective action fixed by demanding agreement with the conical
(and thereby local) notion of curvature. A study of the
discrepancies between the two results may reveal some lessons
about the emergence of locality.

\section{Acknowledgements}
I am particularly grateful to Vijay Balasubramanian for encouragement and critical
comments on the manuscript and to Martin Ro{\v c}ek for an
illuminative correspondence. Discussions with and comments of
Tamaz Brelidze, Peng Gao, Klaus Larjo, Tao Liu, Robert Richter,
Michael Schulz, and Joan Sim{\'o}n are kindly acknowledged. This
work was supported in part by DOE grant DE-FG02-95ER40893.


\begin{thebibliography}{99}

\bibitem{regge}
  T.~Regge, ``General Relativity Without Coordinates,'' Nuovo Cim.
  {\bf 19}, 558 (1961)

\bibitem{doubling}
  D.~H.~Adams,
  ``R-torsion and linking numbers from simplicial abelian gauge theories,''
  [arXiv:hep-th/9612009].
  %%CITATION = HEP-TH 9612009;%%

  D.~H.~Adams,
  ``A doubled discretisation of abelian Chern-Simons theory,''
  Phys.\ Rev.\ Lett.\  {\bf 78}, 4155 (1997)
  [arXiv:hep-th/9704150].
  %%CITATION = HEP-TH 9704150;%%

\bibitem{topbf}
  P.~Mnëv,
  ``Notes on simplicial BF theory,''
  [arXiv:hep-th/0610326].
  %%CITATION = HEP-TH 0610326;%%

\bibitem{chiral}
  V.~de Beauce and S.~Sen,
  ``Chiral Dirac fermions on the lattice using geometric discretisation,''
  [arXiv:hep-th/0305125].
  %%CITATION = HEP-TH 0305125;%%

\bibitem{marsden}
  M.~Desbrun, A.~Hirani, M.~Leok, J.~Marsden,
  ``Discrete Exterior Calculus,''
  [arXiv:math.DG/0508341].

\bibitem{rocek}
  A.~Ko and M.~Ro{\v c}ek,
  ``A gravitational effective action on a finite triangulation,''
  JHEP {\bf 0603}, 021 (2006)
  [arXiv:hep-th/0512293].
  %%CITATION = HEP-TH 0512293;%%

  A.~Ko and M.~Ro{\v c}ek,
   ``A gravitational effective action on a finite triangulation as a discrete
  model of continuous concepts,''
  [arXiv:hep-th/0605022].
  %%CITATION = HEP-TH 0605022;%%

\bibitem{ziliao}
  P.~Di Francesco, P.~Mathieu, D.~S{\' e}n{\' e}chal,
  ``Conformal Field Theory'', Springer-Verlag, New York, 1997.

  M.~Nakahara, ``Geometry, Topology and Physics'', Institute of
  Physics Publishing, Bristol and Philadelphia, 1990.

\bibitem{stanley}
  R.~P.~Stanley, ``Topics in Algebraic Combinatorics,'' Course
  notes for Mathematics 192 (Algebraic Combinatorics), Harvard
  University, Fall 2000.

\bibitem{cubic}
  S.~Sen,
  ``A Cubic Whitney and Further Developments in Geometric Discretisation,''
  [arXiv:hep-th/0307166].
  %%CITATION = HEP-TH 0307166;%%

\bibitem{chung}
  F.~R.~K.~Chung, ``Spectral Graph Theory,'' American Mathematical
  Society, Providence, RI, 1997.

\bibitem{kirchhoff}
  G.~Kirchhoff, ``Über die Auflösung der Gleichungen, auf welche
  man bei der untersuchung der linearen verteilung galvanischer
  Ströme geführt wird,'' Ann. Phys. Chem. {\bf 72}, 497-508 (1847)

\end{thebibliography}
\end{document}